\documentclass[twocolumn,amssymb,12pt]{iopart}
\usepackage{graphicx}
\usepackage{tabularx}
\usepackage{iopams}
\usepackage{ulem}
\usepackage{cite}
\usepackage{xcolor}

\begin{document}
\paper [] {Interplay of structure, magnetism, and superconductivity in Se substituted iron telluride with excess Fe}
\author{Dona Cherian$^1$, S. R\"o\ss{}ler$^2$,
S Wirth$^2$, and Suja Elizabeth$^1$}
\address{$^1$ Department of Physics, Indian
Institute of Science, Bangalore-560012, India}
\address{$^2$ Max Planck Institute for Chemical Physics of Solids,
N\"othnitzer Stra\ss e 40, 01187 Dresden, Germany}
\ead{donacherian@physics.iisc.ernet.in}
\begin{abstract}
We investigated the evolution of the temperature-composition phase diagram of Fe$_{1+y}$Te upon Se substitution. In particular, the effect of Se substitution on the two-step, coupled magneto-structural transition in Fe$_{1+y}$Te single crystals is investigated. To this end, the nominal Fe excess was kept at $y$ = 0.12. For low Se concentrations, the two magneto-structural transitions displayed a tendency to merge. In spite of the high Fe-content, superconductivity emerges for Se concentrations $x\geq$ 0.1. We present a temperature-composition phase diagram to demonstrate the interplay of structure, magnetism, and superconductivity in these ternary Fe-chalcogenides.
\end{abstract}
\maketitle
\section{Introduction}
\indent
In Fe-based chalcogenides and pnictides, many physical properties are strongly influenced by subtle changes in the crystal structure and stoichiometry. In the case of iron chalcogenides, the interstitial Fe plays a crucial role, and the delicate dependence of magnetic ground states on the exact concentration of excess Fe ($y$) in Fe$_{1+y}$Te is well documented \cite{LiPRB2009,BaoPRL2009,RodriguezPRB2011,RosslerPRB2011,ZaliznyakPRB2012, KozPRB2013}. Homogeneous Fe$_{1+y}$Te in the tetragonal structure can be stabilized for $y$ only in the range $0.06\leq y \leq 0.15$ \cite{KozPRB2013}. Within this compositional range, the physical properties vary drastically.   For a narrow range of $0.11\leq y \leq 0.13$, instead of a single first order transition, two transitions occur: an incommensurate antiferromagnetic (AFM) transition at the N\'{e}el temperature $T_{N}$,  which is also associated with a structural change from a tetragonal to an orthorhombic crystal structure.  At lower temperature, the above transition is followed by a first-order transition (identified as  $T_{S}$). At this second phase transition, the crystal symmetry changes to a monoclinic phase \cite{KozPRB2013,CherianJAP2014} and the incommensurate AFM structure becomes commensurate \cite{RodriguezPRB2011}. Thus, both the phase transitions consist of structural as well as magnetic components. The transition at $T_{S}$ is associated with a strong thermal hysteresis. In addition, recent neutron scattering studies have identified a bond-order wave at $T_{S}$ suggesting an electronic origin of this transition \cite{FobesPRL2014}. While the parent compound is antiferromagnetic, substitution of Te by Se induces superconductivity. At ambient pressures, the highest superconducting transition temperature of $T_c \approx$ 15 K and a maximum superconducting volume fraction is observed for about 50 \% Se substitution \cite{YehEPL2008,SalesPRB2009}. While a persistence of the tetragonal structure was observed down to lowest temperatures for superconducting samples with 0.1$\leq x \leq$0.2 \cite{MartinelliPRB2010}, for Fe$_{1.03}$Se$_{0.57}$Te$_{0.43}$ a low temperature orthorhombic structure was found \cite{GrestyPRB2009} at pressure up to about 2~GPa,  and a monoclinic structure above 3~GPa. In spite of many similarities, a considerable difference has been observed between the temperature-composition phase diagrams of Fe pnictides and chalcogenides. In some pnictides, bulk superconductivity evolves once a spin density wave is suppressed, i.e. both phases exclude each other \cite{ZhaoNature2008,LuetkensNature2009,ChuPRB2009}. In others, the two phases coexist in a narrow composition range  where the long-range magnetic ordering takes place at temperatures above the superconducting transition \cite{ChenEPL2009,MartinelliPRL2011,DrewNature2009}. In iron chalcogenides, on the other hand, an intermediate composition regime exists with short-range magnetic ordering which is characterized by charge carrier localization \cite{KhasanovPRB2009,LiuNatureMat2010,HuPRB2013}. The crossing-over from ($\pi,0$) (defined in the crystallographic Fe$_{1+y}$Te lattice) long range order in Fe$_{1+y}$Te into a ($\pi,\pi$) magnetic resonance in substituted superconducting Fe$_{1+y}$Te$_{1-x}$Se$_{x}$ reinforces the view that an intermediate composition regime exists within which short-range magnetic order and superconductivity compete \cite{LiuNatureMat2010,QiyPRL2009,FangPRB2008}. An almost completely superconducting volume fraction is observed when the ($\pi,0$) order is strongly suppressed \cite{LiuNatureMat2010}.

The occurrence of superconductivity in proximity to an antiferromagnetic order suggests a pairing mechanism mediated by spin fluctuations \cite{QiyPRL2009,HanaguriScience2010}. Inelastic neutron scattering studies on superconducting and non-superconducting Fe$_{1+y}$Te$_{1-x}$Se$_{x}$ also revealed spin fluctuations dominated by incommensurate excitations \cite{LumsdenNature2010,ArgyriouPRB2010}. The suppression of long range magnetic order and the emergence of a superconducting transition with increase in Se composition have been investigated by various groups \cite{MartinelliPRB2010,KhasanovPRB2009,HuPRB2013,DongPRB2011,ViennoisJSSC2010}. The temperature composition phase diagram exhibits three regions, as Se composition increased up to 50\%: commensurate AFM order followed by a region where incommensurate AFM order and superconductivity coexist, and bulk superconductivity \cite{KhasanovPRB2009}. In Fe$_{1+y}$Se$_{0.25}$Te$_{0.75}$, for low concentrations of Fe ($y<$0), bulk superconductivity and incommensurate magnetic order coexist, whereas for $y \geq$ 0  bulk superconductivity is suppressed by strong incommensurate magnetic correlations \cite{BendelePRB2010}. On the other hand, Fe$_{1+y}$Te$_{1-x}$Se$_{x}$ with $y$ in the range 0.02 - 0.05 and $x$ $\leq$ 0.075, undergoes a phase transition to an antiferromagnetically ordered phase with a monoclinic structure\cite{MartinelliPRB2010}. However, these previous studies did not address the effect of Se substitution in Fe$_{1+y}$Te with Fe content $y\geq 0.11$, where the temperature dependence of the magnetic and electronic structures becomes more complex. It is also an intriguing question, whether superconductivity still appears with Se substitution in the presence of a large amount of excess Fe. In order to tackle these questions, we investigated single crystals of Fe$_{1+y}$Te$_{1-x}$Se$_{x}$ with $x$  ranging from 0.02 to 0.5 and $y \approx$ 0.12. In the ensuing discussions, the role of composition in altering the magnetic and superconducting properties are outlined.

\section{Experimental}
\indent
Single crystals of Fe$_{1+y}$Te$_{1-x}$Se$_{x}$ (nominal $x$ = 0, 0.02, 0.05, 0.1, 0.15, 0.20, 0.25, 0.4, and 0.5) were grown by a modified horizontal Bridgman method following the same route as outlined in Ref. \cite{RosslerPRB2011}. The nominal concentration of Fe was kept constant at $y$ = 0.12. Powder X-ray diffractograms (XRD) were taken using a Bruker D8 Advanced system after carefully crushing the crystals into a fine powder. Composition analysis was carried out on cleaved samples employing wavelength dispersive spectroscopy with an Electron Probe Micro Analyzer (EPMA), JEOL-JXA-8530F. For bulk measurements, freshly cleaved single crystal surfaces were studied. Specific heat was measured by a relaxation method using a Quantum Design Physical Property Measurement System (PPMS) in the temperature range 2 -- 300~K. In-plane ($ab$-plane) resistivity measurements were carried out in linear four probe geometry utilizing the PPMS. Transport measurements were also conducted in magnetic fields of up to $\mu_0 H = 9$~T, applied parallel to the crystallographic $c$-axis of the crystals. Magnetization was measured using a Quantum Design SQUID magnetometer. For ac susceptibility measurements in the temperature range 4.2~K -- 80~K, a Cryocon ac susceptometer was employed.
%
\section{Crystal structure}
\indent
The XRD patterns recorded at room temperatures on finely powdered single crystals show the phase purity of our as-grown crystals. Within the limits of instrument resolution, the as-grown crystals exhibit a single crystallographic phase without traces of other secondary impurity phases.
 \begin{figure}[!t]
	\centering
	\includegraphics[width=0.55\textwidth]{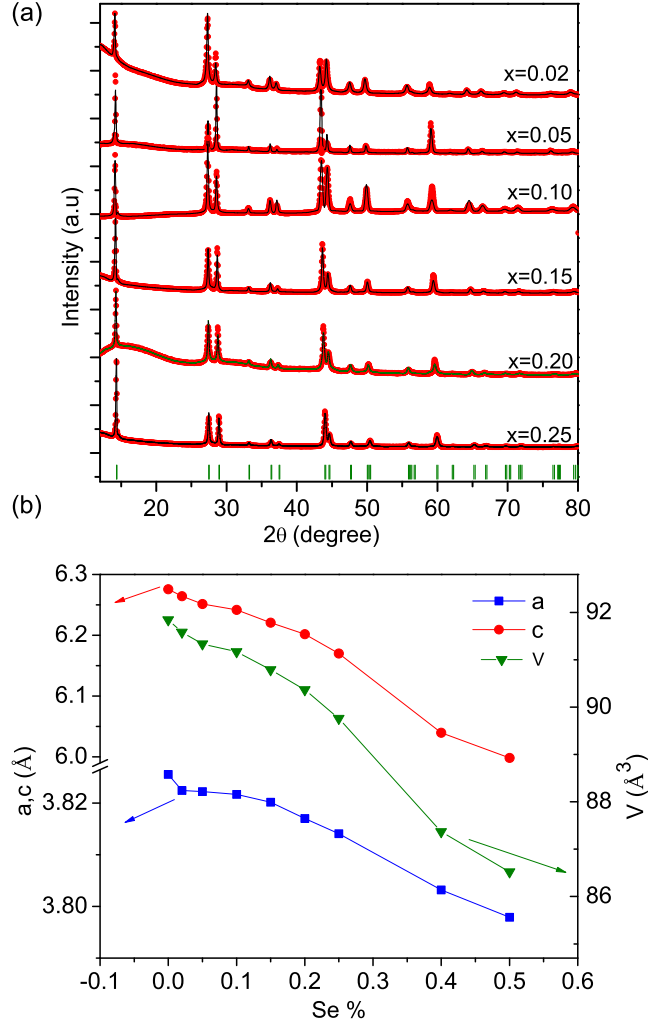}
	\caption{(a) Refined powder XRD data for some exemplary Se substitutions $x$. Red
    symbols represent the experimental data, black lines are the fit results and the green vertical lines indicate the possible Bragg positions. (b) Variation of lattice parameters and unit cell volume with Se substitution.}
	\label{XRD}
 \end{figure}
The effect of substitution on the lattice is studied by XRD. From these data, the lattice parameters are extracted using a Rietveld analysis within the FullProf code and assuming the structure model of an earlier work \cite{RosslerPRB2010}. \Fref{XRD} (a) illustrates the refined powder XRD pattern for the various compositions. The results confirm that the room temperature crystal structure belongs to tetragonal space group $P4/nmm$, typical of Fe chalcogenide superconductors. The variation of the lattice parameters and the unit cell volume as a function of Se substitution is presented in \Fref{XRD} (b). Clearly, a significant contraction of the lattice occurs upon increasing Se substitution $x$.
 \begin{table}[!h]
 \begin{center}
  \caption{\label{EPMA}Composition analysis of Fe$_{1+y}$Te$_{1-x}$Se$_{x}$ samples using EPMA. The estimated values have a standard deviation of 1-2\%.}
  \begin{indented}
  \item[]\begin{tabular}{@{}ll}
   \br
   Nominal & Estimated \\
  \mr
  $y$ = 0.12 & \\
  $x$ = 0    & Fe$_{1.112}$Te\\
  $x$ = 0.02 & Fe$_{1.10}$Te$_{0.98}$Se$_{0.017}$\\
  $x$ = 0.05 & Fe$_{1.09}$Te$_{0.95}$Se$_{0.05}$\\
  $x$ = 0.10 & Fe$_{1.08}$Te$_{0.91}$Se$_{0.09}$ \\
  $x$ = 0.15 & Fe$_{1.10}$Te$_{0.86}$Se$_{0.14}$ \\
  $x$ = 0.20 & Fe$_{1.09}$Te$_{0.81}$Se$_{0.19}$ \\
  $x$ = 0.25 & Fe$_{1.10}$Te$_{0.76}$Se$_{0.24}$ \\
  \br
  \end{tabular}
  \end{indented}
  \end{center}
  \end{table}
A major challenge in these studies is the precise control of the composition of the as-grown samples. For the pristine compound, the composition was determined to Fe$_{1.12}$Te by using synchrotron X-ray diffraction \cite{CherianJAP2014}. All other compositions were estimated by employing the EPMA technique. Upon substitution of Te by Se, a somewhat lower value of excess Fe compared to the nominal $y$ = 0.12 concentration was observed even though the starting compositions were maintained at the same values. The resulting compositions are listed in \Tref{EPMA}. The different samples are labelled with respect to their nominal values of Se for simplicity. Note that even for the pristine sample the Fe excess as estimated by EPMA is slightly lower if compared to the synchrotron XRD result but seemingly larger than in the Se-substituted samples.
\section{ac susceptibility and magnetization}
\begin{figure}[!h]
	\begin{center}
	\includegraphics[width=8cm]{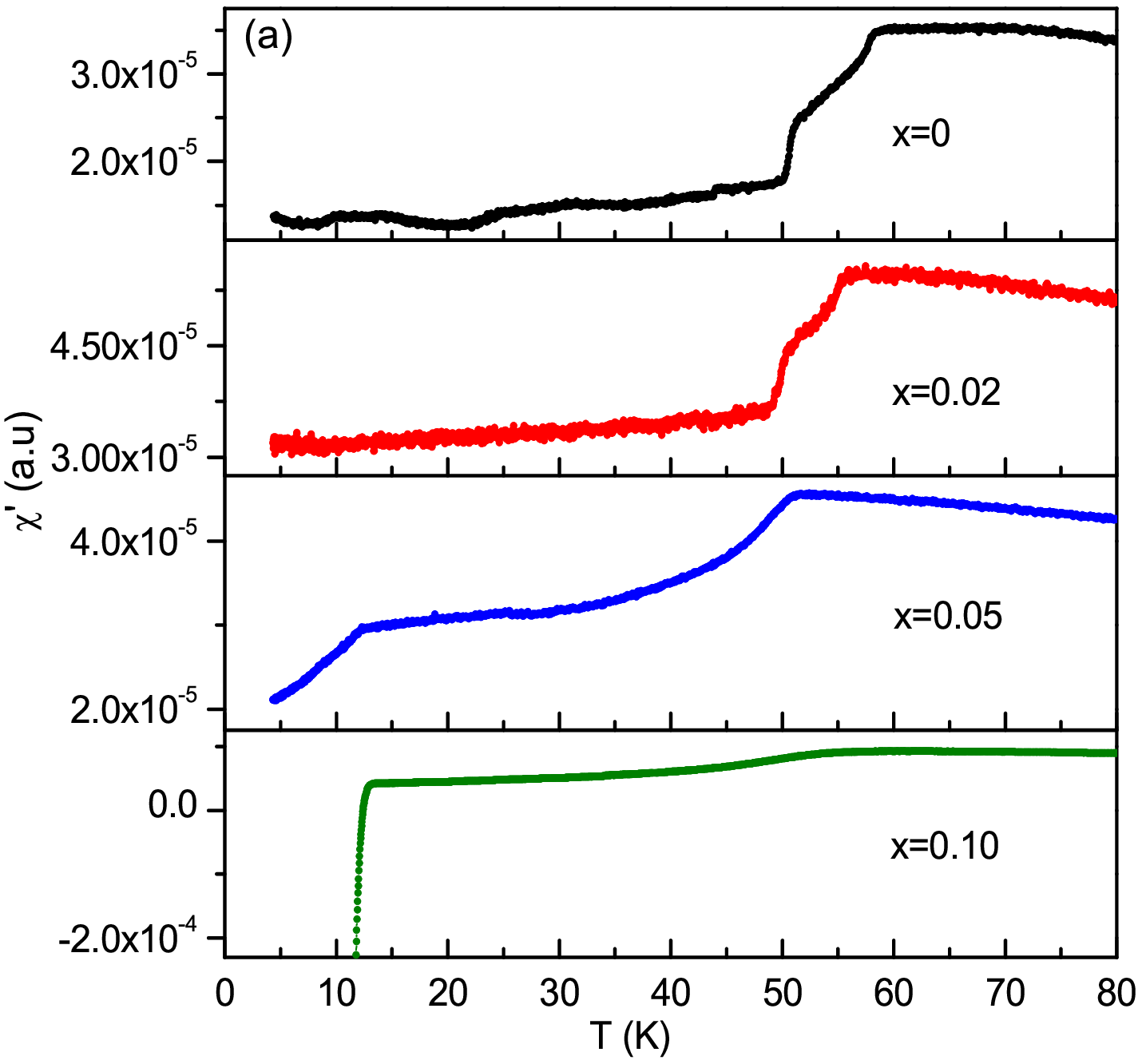}
	\vspace{-0.2cm}
	\includegraphics[width=6.9cm]{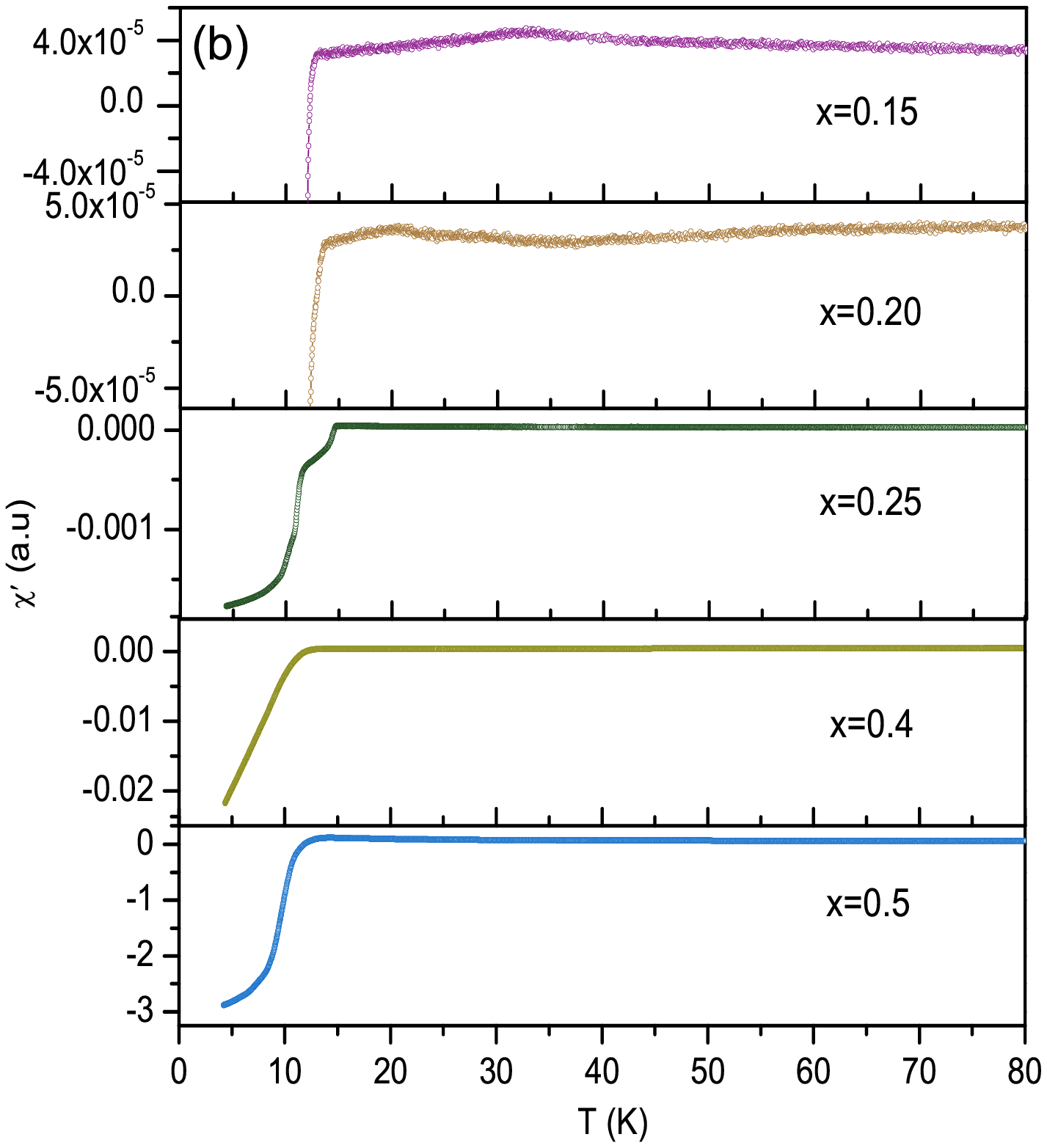}
	\caption{\label{chi}(a) ac $\chi'$ vs $T$ plots of Fe$_{1+y}$Te$_{1-x}$Se$_{x}$ for $x$ = 0 to 0.10 display a change in magnetic transitions. (b) ac $\chi'$ vs T for Se = 0.15 to 0.5. The magnetic transition shifts to low temperature and weakens in the presence of superconductivity.}
	\end{center}
 \end{figure}
The real part of ac susceptibility data are depicted in \Fref{chi}. For $x$ = 0, two slope changes are clearly seen in the transition region, as expected: one corresponds to a second order transition at $T_{N}$= 57~K while the other reflects a first order transition at $T_{S}$ = 46~K \cite{RosslerPRB2011}.
By Se substitution the transitions can be altered. For $x =$ 0.02, the two transitions at $T_N$ and $T_{S}$ can still be resolved although they are slightly shifted to lower temperature, compared to Fe$_{1.12}$Te. Upon increase in Se content to 0.05, $T_N$ falls to 51~K and a down-turn of the susceptibility $\chi$ is observed near 12~K, though the values of $\chi$ remain positive.
Starting with $x =$ 0.10, the superconducting transition is clearly seen at $T_C \approx$ 12~K while the magnetic transition occurs at about 54~K.  Such a presence of a weak magnetic ordering well above the superconducting transition is observed from $x =$ 0.10 onwards. At high Se substitution levels, superconductivity becomes more prominent as indicated by the increasing absolute value of $\chi$ within the superconducting regime. Above $x=0.10$, a weak magnetic transition is still present but shifts to lower temperature with increase in Se content and almost vanishes for $x=0.25$. At still higher Se compositions, $x=$ 0.4 and 0.5, superconductivity dominates \cite{RosslerPRB2010}. The onset of superconductivity and the weakening of magnetic transition in compositions from $x=$ 0.15 to 0.50 are illustrated in \Fref{chi} (b).\\

\begin{figure}[!h]
	\centering
	\includegraphics[width=0.6\textwidth]{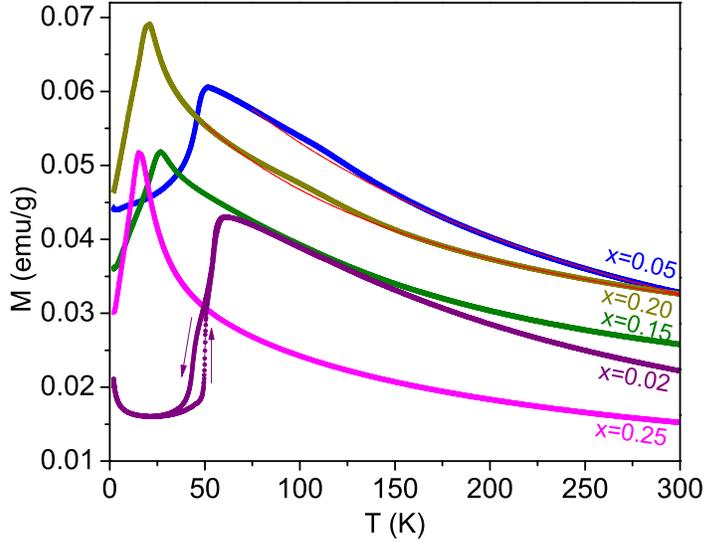}
	\caption{The dc magnetization data obtained at 1000~Oe showing the systematic reduction of $T_N$. The purple up and down arrows indicate the heating and cooling cycle, respectively, for the $x=$ 0.02 sample. The red lines represent the fit.}
	\label{DC}
 \end{figure}
\indent
The dc magnetization of crystals with different composition were studied with magnetic field applied parallel to the crystallographic $ab$ plane, see \Fref{DC}. Here, a relatively large magnetic field of 1000 Oe was used in order to focus on the AFM phase transition. These data display -- more clearly than the ac susceptibility data -- the decrease in $T_N$.

As the Se content increases from $x=$ 0.02 to 0.25, the magnetic transition shifts from 56~K to 16~K. In \Fref{DC}, two slope changes at around  $T_{N}$ = 53.6 K and $T_S$ = 50 K are visible for the sample with $x=$ 0.02. In analogy to Fe$_{1.12}$Te \cite{RosslerPRB2011}, these two successive transitions are attributed to the transition from a tetragonal paramagnetic state to an incommensurate antiferromagnetic phase with orthorhombic structure, followed by an orthorhombic to monoclinic structure with a commensurate antiferromagnetic phase. Only the transition at lower temperature ($T_S$) displays thermal hysteresis, which is very similar to that observed in F$_{1.12}$Te. Fobes \textit{et al.} observed that the hysteretic transition is associated to a development of bicollinear antiferromagnetic order and a ferro-orbital ordering which results in metallic transport behaviour \cite{FobesPRL2014}. Here, by substitution with Se the transition temperatures are altered even though the nature of the transitions appears to be preserved \cite{CherianJAP2014}. While $T_{N}$ is decreased for $x=$ 0.02 compared to the pristine compound, $T_S$ is increased, a behaviour which is also reflected in a reduced temperature span of the transition in the sample $x=$ 0.02. For Se substitution $x >$ 0.02, the thermal hysteresis disappears. The presence of a small amount of Fe$_{3}$O$_{4}$ impurity is detected in two compositions (x=0.05 and x=0.20) as minor anomalies close to 125~K in the dc magnetization measurements. The percentage of impurity is estimated to be less than 0.02\%  by performing a linear fit to the inverse susceptibilities in the paramagnetic region. The red lines in \Fref{DC} represent the magnetization after removing the contribution from Fe$_{3}$O$_{4}$. Both, ac susceptibility and dc magnetization data indicate that in the intermediate composition region, $0.10 \leq x \leq 0.25$, superconductivity emerges from an antiferromagnetically ordered state.

\section{Specific heat}
\indent
A better understanding of the phase transitions is obtained from specific heat ($C_p$) measurements, see Figures 4 (a) and (b). Two peaks are observed for the sample $x=$ 0.02 at $T_{N}$ and $T_{S}$ respectively, as in the case of Fe$_{1.12}$Te.
 \begin{figure}[!h]
	\centering
	\includegraphics[width=0.6\textwidth]{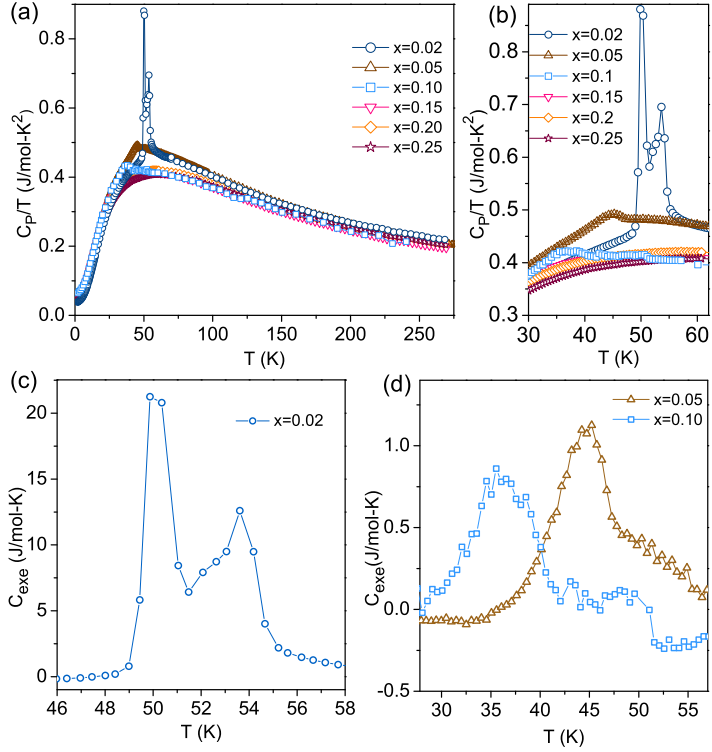}
	\caption{(a) $C_{P}/T$  vs T for different Se substitutions $x$ exhibiting the signature of a phase transitions. (b) Temperature zoom into transition region. (c) Excess specific heat calculated for $x=$ 0.02 showing the split transitions similar to Fe$_{1.12}$Te. (d) Excess specific heat for $x=$ 0.05 and 0.10.}
	\label{Cp}
 \end{figure}
These measurements verify the trend observed above, namely that $T_N$ decreases and $T_S$ increases when $x$ is varied from 0 to 0.02, i.e. the temperature difference between the two transitions is drastically reduced from 11~K to 3.6~K, with $T_{N}$ = 53.6~K and $T_{S}$ = 50~K for sample $x=$ 0.02. This is significant as $T_{N}$ in unsubstituted Fe$_{1+y}$Te is accompanied by a structural transition to an orthorhombic and, at lower temperature, to a monoclinic distortion \cite{RodriguezPRB2011,ZaliznyakPRB2012,KozPRB2013,CherianJAP2014}. In the substituted sample, the shift of the peaks suggests a tendency of the two structural transitions to weaken as well as to merge upon increasing Se content.

The effect of Se substitution on the two transitions is even more pronounced for the $x=$ 0.05 sample. At $T_{N} \approx$ 45~K, the transition in specific heat turns into a broad hump which could either be the result of two weak transitions occurring at very close temperature or of a broad transition. Similarly, for $x$ = 0.1, a weak hump is present near 36~K. The specific heat data of samples 0.02 $\leq x \leq$ 0.10 are analyzed in detail with respect to the weak transitions: The lattice contribution is modelled using a combination of Debye and Einstein models as discussed in \cite{CherianJAP2014}. The model is fitted to our experimental data in the range 2~K $\leq T \leq$ 275~K excluding the transition regions. The lattice ($C_p^{l} = C_p^{Debye} + C_p^{Einstein}$) and electronic ($C_p^{el} = \gamma T$) contributions are subtracted from the total specific heat to obtain the excess specific heat, $C_{exe} = C_p - (C_p^{l} + C_p^{el})$, near the magnetic transition, see Figures 4 (c) and (d). The electronic contribution to the specific heat (represented by $\gamma$ = $C_p^{el} / T$) for each composition is given in Table 2. The sample with $x$ = 0.05 exhibits a peak at around 45~K, followed by a broad shoulder near 49~K. However, in the sample with $x$ = 0.1, a single, broad peak is observed at $\sim$35~K with a small kink close to 50~K. Because of the small magnitude of the peaks, it is difficult to draw conclusions for these substitution levels. The split peaks observed in the pristine compound tend to merge and vanish at low substitution. At higher concentration of Se ($x > $0.1), specific heat data appear not to exhibit any feature corresponding to magnetic ordering likely indicating the absence of long-range magnetic order. These results suggest that at intermediate compositions, a weak magnetic and superconducting phase may exist. Yet, an analysis of the low temperature region of $C_{P}$ provided no hint towards a peak relating to superconductivity.
\section{Resistivity}
To elucidate the superconducting behaviour for intermediate Se content, we examined the electrical transport properties in detail.
 \begin{figure}[!t]
	\centering
	\includegraphics[width=0.6\textwidth]{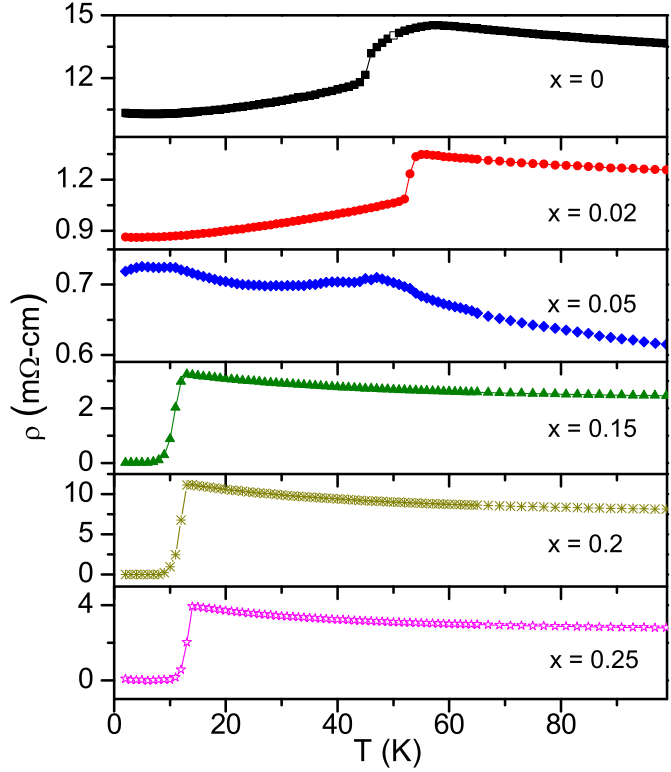}
	\caption{Resistivity vs temperature for $x$=0 to 0.25.}
	\label{RT}
 \end{figure}
The zero field resistivities $\rho$ are plotted in \Fref{RT}. The normal state resistivities of our single crystals exhibit a negative logarithmic dependency for $T_N < T <$ 100~K which is similar to the one observed earlier in Fe$_{1.09}$Se$_{0.5}$Te$_{0.5}$, implying a charge carrier localization \cite{HuPRB2013,RosslerPRB2010,LiuPRB2009}. This is typical for compositions with high Fe concentration. Samples with $x$ = 0 and $x$ = 0.02 show similar trends above and below the magneto-structural transition. The wider transition region in the pristine compound, compared to the 2\% Se-substituted one, is in agreement with the larger temperature difference between $T_{N}$ and $T_{S}$ observed in the former. Below the transition, resistivity follows a $T^2$ dependence in both samples. For $x$ = 0.05, a drastic change in transport behaviour is observed. The hump in $\rho(T)$ just below 50~K corresponds to the magnetic transition as observed in susceptibility and specific heat measurements. At temperatures below the hump, the resistivity does not follow a $T^2$ dependence. Below 10~K, $\rho(T)$ takes a downturn, in agreement with the ac susceptibility measurements. For $x =$ 0.15, superconductivity emerges (at $T^{onset}_{C}$) and $\rho(T)$ drops to zero at $T^{zero}_{C}$. As the Se content increases further, $T^{onset}_{C}$ changes from 12.8~K to 14~K whereas $T^{zero}_{C}$ increases from 7.1~K to 10~K.
 \begin{figure}[!t]
	\centering
	\includegraphics[width=0.65\textwidth]{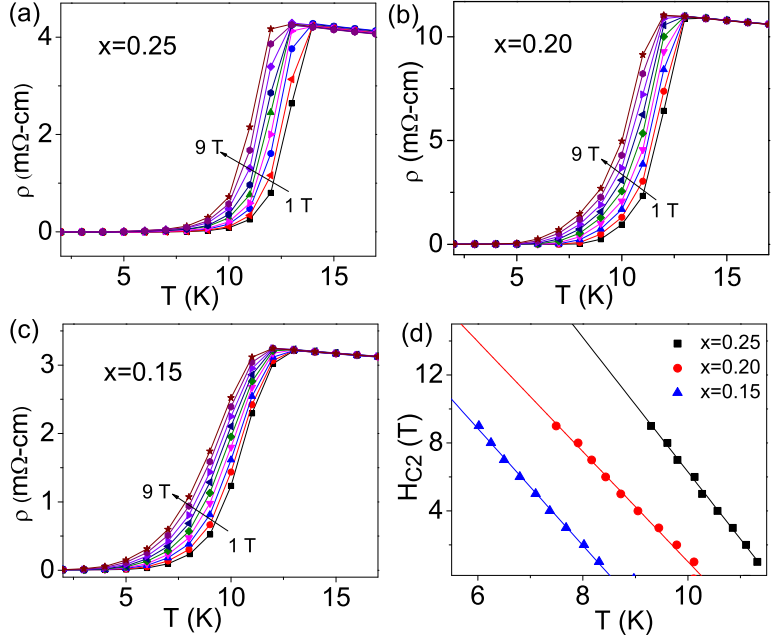}
	\caption{(a) to (c) give the temperature dependent resistivity in external magnetic field for $x$=0.15 to 0.25, demonstrating the shifting of $T_C$ with field. (d) Estimation of $H_{C2}^{zero}$ corresponding to zero resistance. The solid lines represent WHH fit to the data.}
	\label{WHH}
 \end{figure}

We also measured the in-plane resistivity response to applied magnetic fields. The normal state resistivity did not show any significant dependency on $H$. Specifically, the transitions for samples $x=$ 0 and 0.02 remain mostly unaffected (not shown). For samples exhibiting superconductivity, however, $T_{C}$ is influenced by $H$. Figures 6 (a) to (c) provide the resistivity response under magnetic field which shows a decrease of $T_C$ in a similar manner as in type 2 superconductors. A popular approach to analyze the upper critical field $H_{c2}$ is the Werthamer-Helfand-Hohenberg (WHH) formula \cite{WHHPRL1966}. The $H$-$T$ phase diagrams in \Fref{WHH} (d) represent the $H_{c2}$ values at $T^{zero}_{C}$. Applying the WHH theory and assuming a one-band model, the pair breaking field is given by the relation
\begin{eqnarray}
   \centering
   H_{c2}(0) & = &  -0.693 T_{C} \left(\frac{d\,H_{c2}}{d\,T}\right)_{T_{C}}
\end{eqnarray}
where $H_{c2}(0)$ is the zero temperature upper critical field. The calculated values $H_{c2}(0)$ increase with Se substitution, see \Tref{transport}.
\begin{table}[!h]
 \centering
 \caption{\label{transport} Temperatures for onset of superconductivity $T_C^{onset}$ and for zero resistivity $T_C^{zero}$ and zero temperature upper critical fields $H_{c2}(0)$. The electronic contribution to the specific heat in terms of $\gamma$ = $C_p^{el} / T$ is estimated from specific heat measurements.
 }
  \begin{indented}
    \item[]\begin{tabular}{@{}lllllll}
     \mr
  Se\%  & $T^{onset}_{C}$ & $T^{zero}_{C}$ & $ H_{c2}(0)$ & $\gamma$ \\
   &  (K)  & (K)  & (T) & mJ/mol K$^2$ \\
  \mr
  0.02 &  - & - & - & 39.8\\
  0.05 & - & - & - & 48.6\\
  0.10 & - & - & - & 63.4 \\
  0.15 & 12.2 & 7  & 15.04 & 52.6  \\
  0.20 & 13 & 9  & 20.2 & 45   \\
  0.25 & 14 & 11  & 30.1 & 45.4    \\
  \mr
  \end{tabular}
  \end{indented}
\end{table}
\section{Phase diagram}
\begin{figure}[!h]
\centering
\includegraphics[width=0.7\textwidth]{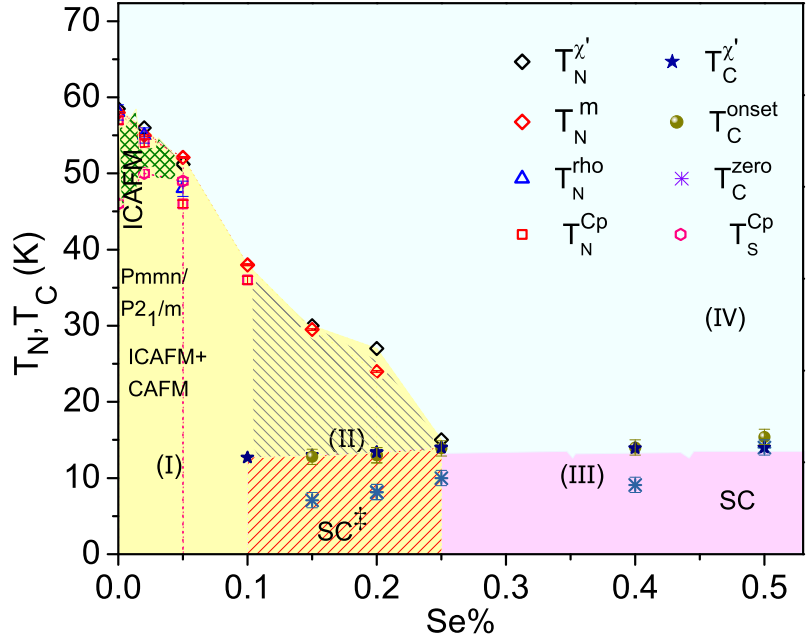}
\caption{\label{PD}Temperature-composition phase diagram. $T_{N}^{\chi '}$, $T_N^m$, $T_{N}^{rho}$ and $T_{N}^{C_p}$ indicate the magnetic transition temperature from ac susceptibility, dc magnetization, resistivity and heat capacity measurements, respectively. $T_{S}^{C_p}$ marks the structural transition. The superconducting transition was obtained from ac susceptibility, $T_{C}^{\chi '}$, and resistivity, $T^{onset}_{C}$ and $T^{zero}_{C}$. Error bars of the estimated transition temperatures are also given along with the respective data points. For further details see Section 7.}
\end{figure}
The overall results and inferences obtained from our single crystals of pristine and substituted Fe$_{1+y}$Te$_{1-x}$Se$_{x}$ are summarized in the $T$-$x$ phase diagram shown in \Fref{PD}. The points in the phase diagram were obtained from  magnetization, specific heat and electronic transport measurements. The phase diagram can be broadly categorized into four regions. Region (\textrm{I}) includes orthorhombic, monoclinic and bicollinear antiferromagnetic phases. The pristine composition (Fe$_{1.12}$Te) undergoes multiple structural transitions at $T_N$ and $T_S$. Similar transitions are observed in crystals with $x$ = 0.02 and 0.05 even though the latter identifies a broad transition in specific heat data. For the pristine Fe$_{1.12}$Te an incommensurate (IC) AFM order was found below $T_N$ which turns into a more complex magnetic order below $T_S$ that contains both commensurate (CAFM) and incommensurate (ICAFM) contributions \cite{KozPRB2013}. We speculate that similar types of magnetic order are present for the sample $x =$ 0.02. Also, for the composition $x =$ 0.05 both commensurate and incommensurate magnetic order might be present. The pink dotted line in the phase diagram represents the boundary where the structural transition vanishes. Region (\textrm{II}) represents the intermediate compositions which exhibit short-range magnetic ordering and superconductivity. The region marked as 'SC$^\ddag$' shows superconductivity which is likely not of bulk nature. Above this, only weak magnetic transitions are observed. Region (\textrm{III}) marks the compositions where magnetic transitions are absent and superconductivity dominates. At temperatures above the transitions, region (\textrm{IV}), the tetragonal, paramagnetic phase prevails.\\
\indent 
Different groups have studied parts of the phase diagram of Fe$_{1+y}$Te$_{1-x}$Se$_{x}$ in detail \cite{MartinelliPRB2010,KhasanovPRB2009,DongPRB2011,ViennoisJSSC2010,BendelePRB2010}. These studies mainly investigated the samples with low amount of excess Fe. Our investigations focus on the effect of Se substitution on the two successive phase transitions observed in Fe$_{1.12}$Te. Apparently, the two transitions seem to merge with increasing substitution of Se. This effect can be seen in the region of the phase diagram represented by ICAFM. With further increase in Se composition, the antiferromagnetic transition is drastically suppressed. The behaviour of the transitions of the $x=$ 0.02 sample is in close resemblance with that observed in Fe$_{1.12}$Te as seen from bulk measurements. A comparison with Fe$_{1.12}$Te which has a mixed crystallographic phase below $T_S$ hints at the possibility of a mixed phase in low Se substituted compositions below $T_S$. Thus, the low Se composition region in the phase diagram can be divided into two parts, one with ICAFM order and orthorhombic phase and the second one with a mixed phase. Our results along with previous reports suggest that with the weakening of long-range magnetic ordering, the structural transition is suppressed and the tetragonal symmetry is preserved \cite{MartinelliPRB2010}. Apparently, even with a very low concentration of Fe, bulk superconductivity is observed only for compositions with $x > $ 0.3 \cite{LiuNatureMat2010,HuPRB2013}. Our phase diagram provides a region where multiple structural/magnetic phase transitions are present which are eventually suppressed by Se substitution. Although superconductivity emerges with higher amount of excess Fe,  100\% superconducting volume fraction is not attained for Fe-rich compositions. 

\section{Conclusion}
\indent
The magnetic and superconducting properties of Fe$_{1.12}$Te$_{1-x}$Se$_{x}$ ($0 \leq x \leq 0.5$) are explored by several measurement techniques on identical samples. By intentionally choosing a high Fe concentration in these Se substituted compositions, the multiple structural transitions are altered. The results of our measurements allow us to construct a comprehensive temperature-Se content phase diagram. The magnetic and structural transitions shift, and finally disappear, upon gradually increasing the Se content. With the disappearance of long range ordering, the structural transition is suppressed thus preserving a tetragonal symmetry. This is in good agreement with previous studies in which the low temperature structure of intermediate compositions was found to remain tetragonal. The intermediate composition region (0.1 $\leq x \leq$ 0.25) is studied in detail, and our investigations demonstrate the presence of short range magnetic fluctuations above $T_C$. Superconductivity in this composition range is likely not of bulk nature. The results clearly corroborate the role of excess Fe in controlling the magnetic and superconducting properties of Fe$_{1+y}$Te$_{1-x}$Se$_{x}$.

\ack{The funding from the Department of Science and Technology (DST) and the Deutscher Akademischer Austauschdienst (DAAD) is greatly acknowledged.
SR is supported by the Deutsche Forschungsgemeinschaft (DFG) within the Schwerpunktprogramm SPP1458. }
\section*{References}
 \providecommand{\newblock}{}

\end{document}